\documentstyle[floats,twocolumn,aps,psfig,prl]{revtex}
\begin{document}
\draft
\twocolumn[\hsize\textwidth\columnwidth\hsize\csname @twocolumnfalse\endcsname
\title{Critical behavior of the conductivity of Si:P at the metal-insulator
transition under uniaxial stress}
\author{S. Waffenschmidt, C. Pfleiderer, H. v. L\"ohneysen}
\address{Physikalisches Institut, Universit\"at Karlsruhe,
D-76128 Karlsruhe, Germany}
\date{\today}
\maketitle

\begin{abstract}
We report new measurements of the electrical conductivity $\sigma$ of the canonical 
three-dimensional metal-insulator system Si:P under uniaxial stress $S$.
The 
zero-temperature extrapolation of $\sigma(S,T
\rightarrow 0) \sim \mid S - S_c \mid^{\mu}$ shows an unprecidentedly sharp
onset of finite conductivity at $S_c$ with an exponent $\mu = 1$. The value 
of $\mu$ differs significantly from that of earlier stress-tuning results.
Our data show dynamical $\sigma(S,T)$ scaling on both metallic and insulating
sides , viz. $\sigma(S,T) =  \sigma_c(T) \cdot {\cal F}(\mid S - S_c \mid/T^y)$
where $\sigma_c(T)$ is the conductivity at the critical stress $S_c$.  We find $y = 1/z\nu$ = 0.34 where $\nu$
is the correlation-length exponent and $z$ the dynamic critical exponent. 
\end{abstract}
\vskip2pc]

Quantum phase transitions have become of steadily increasing interest in
recent years \cite{sondhi97}. These continuous transitions ideally occur
at temperature $T = 0$ where quantum fluctuations play the role corresponding
to thermal fluctuations in classical phase transitions. In particular, 
certain types of 
metal-insulator transitions (MIT) such as localization transitions have been
studied extensively. Experimentally, the MIT may be driven by an external 
parameter $t$ such as carrier concentration $N$, uniaxial stress $S$, or
electric or magnetic fields. Generally, electron localization might arise 
from disorder (Anderson transition) or from 
electron-electron (e-e) interactions (Mott-Hubbard transition) \cite{belitz94}. 
In Nature, these two features go hand in hand.
For instance, the disorder-induced MIT occurring as a function of doping in
three-dimensional ($d$ = 3) semiconductors where the disorder 
stems from the statistical distribution of dopant atoms in the crystalline 
host, bears signatures of e-e interactions as evidenced from 
the transport properties in both metallic \cite{rosenbaum81}
and insulating regimes \cite{liu96}. This makes a theoretical
treatment of the critical behavior of a MIT exceedingly difficult. 
Even for purely disorder-induced transitions, the critical
behavior of the zero-temperature dc conductivity, $\sigma(0) \sim \mid t - t_c
\mid^{\mu}$ where $t_c$ is the critical value of $t$, is not well understood.
Theoretically, $\mu$ is usually inferred from the correlation-length 
critical exponent $\nu$ via Wegner scaling $\mu = \nu (d-2)$. Numerical 
values of $\nu$ range between 1.3 and 1.6 \cite{kramer90,ohtsuki99}.

Experimentally, it has long been suggested that the critical behavior of the 
conductivity falls into two classes: $\mu \approx 0.5$ for uncompensated 
semiconductors and $\mu \approx 1$ for compensated semiconductors and 
amorphous metals \cite{thomas}. However, there appears to be no clear 
physical distinction between these materials
that would justify different universality classes. While many
different materials were reported to show $\mu \approx 1$, the exponent
$\mu \approx 0.5$ was largely based on the very elegant 
experiments by Paalanen and coworkers \cite{paalanen82,thomas83,rosenbaum83},
where uniaxial stress was used to drive
an initially insulating uncompensated Si:P sample metallic. This allows to 
fine-tune the MIT since the stress can be changed continuously at low $T$ thus eliminating 
geometry errors incurring when different samples are employed in concentration
tuning the MIT. 

As always when dealing with critical phenomena, the range of critical 
behavior is a source of controversy. A few years ago we suggested \cite{stupp93}
to limit the critical concentration region in doped semiconductors on the 
metallic side of the MIT to samples where $\sigma(T)$ actually decreases 
with decreasing $T$, i.e. the sample becomes less conducting when approaching
the MIT. In doped semiconductors, $\sigma(T)$ is nearly independent 
of $T$ at the crossover concentration $N_{cr}$, with a value $\sigma_{cr}$
of a few times $10\,\Omega^{-1} {\rm cm}^{-1}$, e.g. $\sigma_{cr} \approx
40\,\Omega^{-1} {\rm cm}^{-1}$ in Si:P. $\sigma (T)$ exhibits a negative 
temperature coefficient above $N_{cr}$ which is explained in terms of 
e-e interactions \cite{rosenbaum81}. Typically the critical region 
$N_c < N < N_{cr}$ is within 10$\%$ or less of the critical concentration $N_c$. 
This eliminates a large number of studies purporting to show
$\mu = 0.5$ where actually only a few samples in the critical regime were 
investigated. Even the recent study on transmutation-doped Ge:Ga where 
$\mu = 0.5$ was suggested, presents only three metallic samples in 
the critical region below $\sigma_{cr} \approx 10\,\Omega^{-1}{\rm cm}^{-1}$
\cite{itoh96}. An earlier study of a large number of Si:P samples 
showed that the 
conductivity exponent $\mu$ changed from $\mu = 0.64$ for $N > N_{cr}
\approx 1.1\,N_c$ to
1.3 for $N_c < N < N_{cr}$ \cite{stupp93}. On the other hand,  
sample inhomogeneities might affect  the behavior very close to $N_c$. For
this reason, data for stress-tuning with stress close to the critical value
were discarded in the earlier study, leading to $\mu = 0.5$ 
\cite{paalanen82,thomas83,rosenbaum94}. It is therefore absolutely
necessary to perform additional stress-tuning experiments on Si:P with finely 
tuned stress values including data on the insulating side to check for the critical behavior.

The notion of a quantum phase transition allows a second important 
aspect to be addressed, namely the interdependence of static and dynamic 
behavior. The dynamics is reflected in the finite-temperature behavior of
critical quantities. Concerning the MIT in heavily-doped semiconductors,
this point has not received much attention from the experimental side. 
A first attempt was made \cite{stupp94} using the scaling function \cite{belitz94}
\begin{equation}
\sigma(t,T) = (t - t_c)^{\mu} {\cal F} (T/(t - t_c)^{z \nu})
\end{equation}
where $z$ is the dynamic critical exponent. This relation is often referred 
to as dynamic scaling. Approximate dynamic scaling was observed for 
Si:P on the metallic side of the MIT with $t = N$, yielding $\mu = 1.3$
and $z = 2.4$ \cite{stupp94}. On the other hand, the stress-tuning data 
\cite{thomas83} did not obey scaling \cite{belitz94}. Very recently, 
Bogdanovich et al. \cite{bogdanovich99} demonstrated that conductivity 
data for Si:B under uniaxial stress obey very
nicely the dynamic scaling  on both metallic and insulating sides,
yielding $\mu = 1.6$ and $z = 2$, while concentration tuning of $\sigma(0)$
on the same
system had suggested $\mu = 0.63$ \cite{dai91}. This large difference is
not understood at present. In this situation, an examination
of possible dynamic scaling of the canonical metal-insulator system Si:P
appears of utmost importance in order to resolve the question of critical behavior
and to appraise the possibly strongly different roles of stress and concentration
in tuning the MIT.

In this paper, we report on stress tuning of the MIT of Si:P by measuring
the electrical  conductivity down to 15\,mK. We find by extrapolating to 
$T = 0$ an unprecidently sharp onset of $\sigma (t,0)$ which allows to 
unambiguously extract
$\mu \approx 1$. In addition, dynamic scaling yielding $z \sim 3$ is
found. The value of $\mu$ is in reasonable agreement with that derived from 
concentration tuning. We further demonstrate that stress tuning and concentration
tuning lead to very different $T$ dependencies of $\sigma$.

The samples were taken from the same Si:P crystals which have been employed
previously \cite{stupp93}. Here we report on investigations on two crystals
with $N = 3.21$ and 3.43 $\cdot 10^{18}{\rm cm}^{-3}$, just below the critical
concentration $N_c = 3.52 \cdot 10^{18}{\rm cm}^{-3}$ as determined 
\cite{stupp93} for our samples. Similarly grown samples with an even higher
concentration ($N \approx 7 \cdot 10^{19}{\rm cm}^{-3}$) showed no sign of 
P clustering as investigated with scanning tunneling microscopy 
\cite{trappmann97}. The samples were cut to a size of $\sim$ 15\,x\,0.8\,x\,0.9\,mm$^3$ 
and contacted with four Au leads by spark welding, 
with the voltage leads $\sim$ 6\,mm apart. The sample was mounted in a 
$^4$He-activated uniaxial pressure cell equipped with a piezoelectric force 
sensor. The stress was applied along the [100] direction which was the most
elongated dimension of the sample. The stress was determined from the
ratio of the area of the cell base plate and the sample cross section. 
Calibration of the cell showed a linear increase of force with pressure 
applied at room temperature to gazeous He, with no hysteresis between 
increasing and decreasing pressure. The cell, incorporating a thermal
shield, was tightly screwed to the mixing chamber of a dilution
refrigerator. During one run a thermometer was attached to the sample showing 
that temperature deviations to the main thermometer directly mounted at the 
mixing chamber were less than 0.5\,mK  at the lowest measuring temperature of 
15\,mK. The conductance was measured with a LR 700 resistance bridge at
16\,Hz.

\begin{figure}[h]
	\centerline{\psfig{file=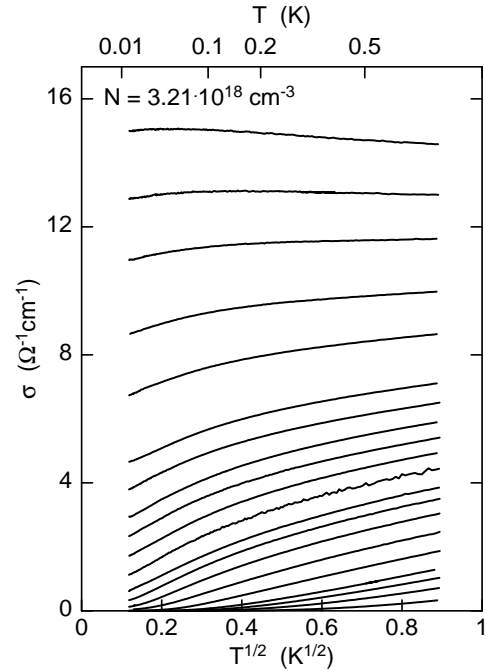,width=7cm,clip=true}}
	\caption{Conductivity $\sigma$ of a Si:P sample with P concentration $N =
3.21 \cdot 10^{18}{\rm cm}^{-3}$ versus $\surd T$ for several values 
of uniaxial stress. 
From top to bottom: $S$ = 3.05, 2.78, 2.57, 2.34, 2.17, 2.00, 1.94, 1.87, 1.82,
1.77, 1.72, 1.66, 1.61, 1.56, 1.50, 1.41, 1.33, 1.26, 1.18, 1.00\,kbar.
Solid lines are connecting the very finely spaced individual data points.}
\end{figure}

Fig.\,1 shows the electrical conductivity $\sigma(T)$ of sample\,1 
($N = 3.21 \cdot 10^{18}{\rm cm}^{-3}$) for uniaxial pressures between
1 and 3.05 kbar. The data are plotted vs. $\surd T$ which is the $T$
dependence expected due to e-e interactions
and indeed observed well above the MIT, $\sigma (T) = 
\sigma_0 + m \surd T$ with $m < 0$ 
\cite{rosenbaum81}. The smooth curves are in
fact polygons connecting adjacent
data points (see Fig.\,2a for a  set of actual data points). Under uniaxial
stress between 1 and 2.57\,kbar the $\sigma(T)$ curves evolve 
smoothly from insulating to metallic behavior with 
$m > 0$, and $\sigma (T)$ becomes nearly independent of $T$ with a value
$\sigma_{cr} \approx 12\,\Omega^{-1}{\rm cm}^{-1}$ at $\sim$ 2.7\,kbar. For
larger stress $\sigma (T)$ passes over a shallow maximum signaling the crossover
to $m < 0$, as observed with concentration tuning \cite{blaschette96}.
It is interesting to note that $\sigma_{cr} (S) \approx 0.3\,\sigma_{cr}(N)$,
thus severely limiting the critical region. Our data do not exhibit the 
precipitous
drop of $\sigma (T)$ below $\sim$ 40\,mK for pressures closest to the MIT,
in distinction to the earlier stress-tuning work on Si:P extending to 3\,mK
\cite{paalanen82,thomas83}. Instead, our $\sigma (T)$ data exhibit a $T$ 
dependence that varies only gently with stress.

\begin{figure}[h]
	\centerline{\psfig{file=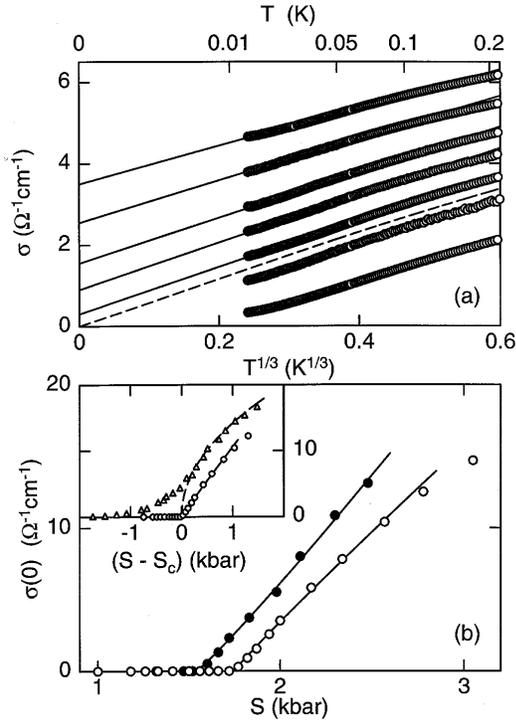,width=7cm,clip=true}}
	\caption{(a) Low-temperature data of $\sigma$ of Fig.\,1 in the immediate
vicinity of the metal-insulator transition plotted against $T^{1/3}$. 
Dashed line indicates the conductivity at the critical stress (see text). 
(b) Extrapolated conductivity 
$\sigma (0)$ for $T \rightarrow 0$ versus uniaxial stress $S$ for two P 
concentrations $N = 3.21$ and $3.43 \cdot 10^{18}{\rm cm}^{-3}$ (closed and
open circles, respectively). The inset shows 
earlier $\sigma (0)$ versus $S - S_c$ data (triangles) from Ref.\,[8] in 
comparison to our data for sample 1 (circles).}
\end{figure}

Closer inspection shows that the data near the MIT are actually better
described by a $T^{1/3}$ dependence for low $T$ as can be seen from
Fig.\,2a for a few selected pressures in the immediate vicinity of the MIT.  
$\sigma (0)$ obtained from the $T^{1/3}$ extrapolation to $T = 0$ is shown 
in Fig.\,2b,  together with data for sample 2 ($N = 3.43 \cdot 
10^{18}{\rm cm}^{-3}$) (see Fig.\,3 for $\sigma (T)$
of this sample for a few representative uniaxial pressures).
$\sigma (0)$  is plotted linearly vs.  $S$, yielding $S_c$ = 1.75\,kbar
for sample\,1 and 1.54\,kbar for sample\,2. Note that 
the critical stress $S_c$ is quite well defined, as $\sigma (0)$ breaks
away roughly linearly from zero within less than 0.1\,kbar. Applying our
criterion for the critical region, the analysis should be limited to data 
with $\sigma < \sigma_{cr} \approx 12\,\Omega^{-1}{\rm cm}^{-1}$. 
In this range
the critical exponent $\mu$ is 0.96 and 1.09 for sample 1 and 2, respectively.
$\mu \approx 1$ are found also when the more conventional $\surd T$ 
extrapolation is employed as can be inferred from Fig.\,1. This behavior 
contrasts with the earlier stress-tuning data \cite{paalanen82} reproduced 
in the inset of Fig.\,2b, where 
appreciable rounding close to $N_c$ is visible as compared to our samples
when plotted against $S - S_c$ (see also \cite{rosenbaum94}). However, those 
$\sigma (0)$ data between 4 and $16\,\Omega^{-1}{\rm cm}^{-1}$ are compatible
with linear dependence on uniaxial stress.

\begin{figure}
	\centerline{\psfig{file=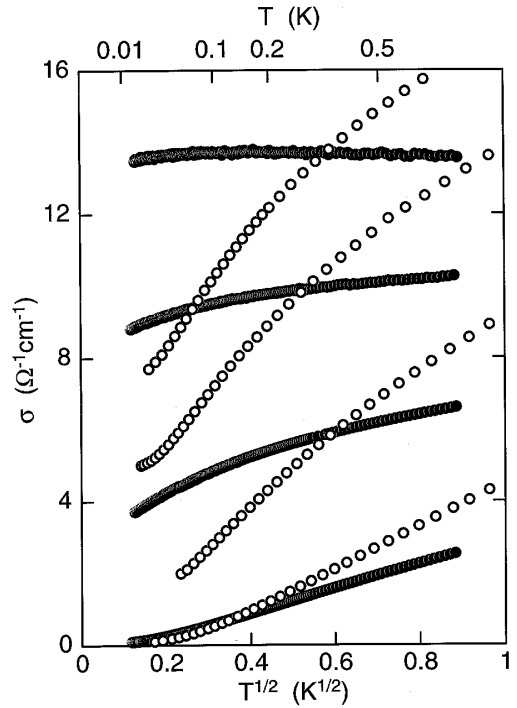,width=7cm,clip=true}}
	\caption[] {Comparison of the concentration dependence of $\sigma (T)$ of 
Si:P (open symbols, from top to bottom: $N = 3.60, 3.56, 3.50, 
3.38 \cdot 10^{18}{\rm cm}^{-3}$, from Ref.\,[11]) and stress dependence of 
$\sigma (T)$ (closed symbols, $N = 3.43  \cdot 10^{18}{\rm cm}^{-3}$, from top to
bottom: $S$ = 2.48, 2.11, 1.72, 1.32\,kbar).}
	\end{figure}

Fig.\,3 shows $\sigma (T)$ of sample\,2 for a range of selected uniaxial 
pressures,
again applied along [100]. The overall behavior is very similar to that
of sample\,1. The fact that $\sigma_{cr}$ is the same for both samples
is nevertheless surprising given the difference in $S_c$. It has been 
suggested that tuning with $S$ or $N$ should yield  the same
critical exponents \cite{paalanen82,thomas83,rosenbaum83,bhatt82}.
The decrease of $N_c$ with uniaxial
stress is attributed to the admixture of the more extended 1$s$(E) and 
1$s$(T$_2$) excited states to the 1$s$(A$_1$) groundstate
of the valley-orbit split sixfold donor 1$s$ multiplet \cite{bhatt82}. 
However, comparison of $\sigma (T)$ for various $S$ and $N$
(Fig.\,3) reveals that stress and concentration
tuning lead to strikingly different $T$ dependences of the conductivity in 
the vicinity of the MIT. As the exact origin of the $\sigma (T)$ behavior
close to the MIT is unknown, we cannot offer an explanation for the different
behavior which, of course, must arise from the  change of donor wave functions
under uniaxial stress. In this respect, experiments on similar samples for 
stress applied along different directions leading to different types
of mixing among the states of the 1$s$ multiplet will be helpful. The fact 
that stress was applied to different directions in the previous and present 
study, i.e. [12$\bar3$] and [100], respectively, may well be one  reason for 
the different behavior of $\sigma (T)$.

We finally turn to the scaling behavior of $\sigma$ at finite temperatures
using the data of sample 1. 
We employ the scaling relation \cite{bogdanovich99}
\begin{equation}
\sigma(t,T) = \sigma_c(T) {\cal F}^{'} ((t - t_c)/T^y)
\end{equation}
where $\sigma_c (T) = \sigma(t_c, T)$ is the conductivity at the critical
value $t_c$ of the parameter $t$ driving the MIT. This scaling relation is 
equivalent
to Eq.\,(1), both are derived from the general scaling relation
\begin{equation}
\sigma (t,T) = b^{-(d-2)} {\cal F}^{''} ((t - t_c) b^{1/\nu}, b^z T)
\end{equation} 
where $b$ is a scaling parameter. If the leading term to $\sigma_c(T)$ is 
proportional to $T^x$, one obtains $x = \mu /\nu z$ and $y = 1/ \nu z$
from a scaling plot. Fig.\,1 and 2a show that $\sigma$ for $S$ close to 
$S_c$ does not exhibit a simple power-law $T$ dependence over
the whole $T$ range investigated. We therefore determine $\sigma_c(T)$ self-consistently
in the  following manner. The critical stress for sample 1 is taken from
the above analysis as $S_c = 1.75$\,kbar. In order to obtain $\sigma_c(T)$,
we interpolate linearly between the two $\sigma (T)$ curves for $S=1.72$ and 
1.77\,kbar. The resultant $\sigma_c(T)$ is then fitted by the function 
$\sigma_c(T) = a T^x(1+ dT^w)$ with $a = 6.01\,\Omega^{-1}{\rm cm}^{-1},
x = 0.34, d = -0.202, w = 0.863$, and $T$ is expressed in K. Here the $dT^w$ 
term presents a correction to the critical dynamics.
This $\sigma_c(T)$ curve is shown in a dashed line in Fig.\,2a. All $\sigma (S,T)$ 
curves with 1.00\,kbar $< S <$ 2.34\,kbar up to 800\,mK are then used for 
the scaling analysis according to Eq.(2). 
The same procedure was repeated for other choices of $\sigma_c(T)$ between 
the two measured $\sigma (T)$ curves embracing the critical stress with 
clearly less satisfactory results.

\begin{figure}[h]
	\centerline{\psfig{file=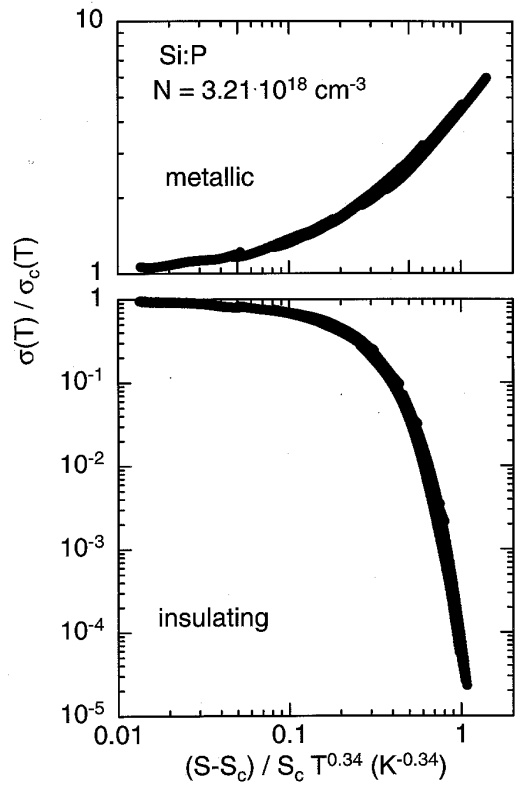,width=7cm,clip=true}}
	\caption{Scaling plot of $\sigma / \sigma_c$ vs. $\mid S - S_c \mid 
/ S_c T^y$ for sample 1, with $S_c = 1.75$\,kbar and $y = 0.34$.}
	\end{figure}

Fig.\,4 shows the resulting scaling plot of $\sigma (S,T)/\sigma_c(T)$
vs. $\mid S - S_c \mid / S_c T^y$. The data are seen to collapse on a 
single branch each for the metallic and insulating side, respectively. The
best scaling, as shown, is achieved for $y = 1/z \nu = 0.34$. Together with 
$\mu = 1.0$ as obtained from Fig.\,1 and assuming Wegner scaling $\nu = \mu$
for $d = 3$, we find $z = 2.94$, which is  indeed consistent with
$\sigma_c \sim T^{1/z} \sim T^{1/3}$ for  $T \rightarrow 0$ (see Fig.\,2a). 
Alternatively, we may use Eq.(1) plotting $\sigma (S,T) / \mid S - S_c \mid^{\mu}$
vs. $T / \mid S - S_c \mid^{z\nu}$  (not shown) with the three parameters $S_c, \mu = \nu$ and
$z$. The best data collapse is found
for $\mu = 1.0 \pm 0.1$ and $z = 2.94 \pm 0.3$, in very good agreement with the values 
obtained from Fig.\,4. Additionally, we note the broad consistence with the
earlier concentration tuning data where $\mu = 1.3$ and $z = 2.4$ was inferred
\cite{stupp94}. We estimate the error of our combined analysis of the
present stress-tuned data to 10\% for
$\mu$ and $z$. The critical stress is determined with a relative 
accuracy to better than 0.1\,kbar. It is important to note that either
$\sigma(0)$ scaling (Fig.\,2b)  or dynamic scaling (Fig.\,4) when taken by
itself, may lead to a rather large error in $\mu$ and/or $z$, just because
of the ambiguity of determining the critical region. However, the consistent
determination of exponents from the combined scaling lends confidence to the 
values reported here.

The above procedure to determine the conductivity at the critical stress 
is necessary because $\sigma_c$  does not obey a simple
power-law $T$ dependence over the whole $T$ range. Above 100\,mK the 
correction term $dT^w$ (with $d < 0$)
comes into play. This is at variance with Si:B where
$\sigma_c \sim T^{1/2}$ was observed in the whole range from 60 to 800\,mK
\cite{bogdanovich99}. On the other hand, a $T^{1/3}$ dependence of $\sigma$ 
in the vicinity
of the MIT has been reported for transmutation-doped Ge:Ga over a large $T$
range \cite{itoh96}. Certainly, the finite-$T$ behavior near the quantum critical
point needs closer theoretical scrutiny, in particular since the dynamic
scaling is observed up to 800\,mK when the correction term to
$\sigma_c(T)$ is included. We remark that a simple 
algebraic $T$ dependence $\sigma_c = aT^x$ which yields good 
dynamic scaling for Si:B \cite{bogdanovich99}, clearly leads to 
less satisfactory scaling in Si:P for any choice of $x$.

In conclusion, we have demonstrated dynamic scaling of stress-tuned Si:P at 
the metal-insulator transition. The conductivity exponent $\mu \approx 1$ 
is close to the 
exponents derived earlier from concentration tuning. However, upon application
of stress, the critical range is narrowed to conductivities below
12\,$\Omega^{-1}{\rm cm}^{-1}$. Therefore, it is the absence of appreciable
rounding effects in our samples  close to the MIT that allows us to determine
$\mu \approx 1$ reliably, thus resolving the conductivity exponent puzzle.
The temperature dependence of the conductivity starting from the same 
$\sigma (0)$ value for $T = 0$ is distinctly different for samples under zero
stress and under stress. It is predicted that in the region between 15 and 
40\,$\Omega^{-1}{\rm cm}^{-1}$ initially insulating stress-tuned samples
will exhibit a negative slope of $\sigma (T)$, while samples under zero stress
in this range are known to have a positive $\sigma (T)$. In view
of these differences away from the quantum critical point, the similarity
of asymptotic dynamic scaling behavior is particularly noteworthy. A 
more detailed theoretical treatment which may eventually
also account for the effective exponent $\mu \approx 0.5$ for samples 
above the crossover conductivity $\sigma_{cr}$ is highly desirable.\\

We thank W. Zulehner, Wacker Siltronic AG, Burghausen, Germany for the samples, and M. P.
Sarachik and P. W\"olfle for useful discussions. This work was supported by 
the Deutsche Forschungsgemeinschaft through SFB 195.

\end{document}